# The Medical Algorithms Project

M Sriram Iyengar, PhD, Asst. Professor, School of Health Information Systems, UTHSCH. Email: m.sriram.iyengar@uth.tmc.edu, 281 793 4733, John R Svirbely, MD, TriHealth, Cincinnati, Ohio

**Abstract**:
*The Medical Algorithms Project, a web-based resource located at www.medal.org, is the world's largest collection of medical-related spreadsheets, consisting of over 13,500 Excel spreadsheets each encoding a medical algorithm from 45 different areas of medical practice. This free resource is in use worldwide with over 106,000 registered users as of March 1, 2009.*

## *Introduction*

A *medical algorithm* is any formula, score, scale, diagram, or computational technique that is useful in healthcare. Examples of medical algorithms include the Glasgow coma scale, TNM stage in oncology, and predictive risk factors for heart disease. Medical algorithms comprise a technology for medical decision support, with the potential to decrease time demands on clinicians, to support evidence-based medicine, to reduce errors, and to help increase the quality while decreasing the cost of care. We estimate that the biomedical peer-reviewed literature contains over 250,000 medical algorithms across all the specialties and sub-specialties of medicine. However, these algorithms are underutilized by busy clinicians because it takes time to find, to understand and to adapt them to a specific clinical situation. Our hope has been that they would be used more often by clinicians and medical researchers if they were readily accessible in a digital format.

The medical algorithms project (www.medal.org) was developed to enhance the availability and accessibility of medical algorithms to medical practitioners such as nurses, physicians, pharmacists, biomedical researchers and others in the healthcare field. This resource contains descriptions and executable versions for over 13,500 medical algorithms in 45 different subject areas (Fig. 1). All of the algorithms are implemented as Microsoft Excel spreadsheets; in addition, approximately 400 spreadsheets are implemented in PHP.

In the following we describe this resource further.

Figure 1: Fragment of Contents page at www.medal.org

## *History*

The medical algorithms project was conceived and developed by John Svirbely, MD, Trihealth Hospitals, Cincinnati, and Sriram Iyengar, PhD, School of Health Information Sciences (SHIS), University of Texas, Houston. They observed that large numbers of computational techniques such as scales and scores were being presented in the biomedical journals and books. These medical algorithms can potentially aid the practice of healthcare since they are derived from experimental or observational data. The numerous benefits for medical practice include support for medical decision-making, increased accuracy of diagnosis and prognosis, use of evidence-based criteria rather subjective judgments, increased scope of physician practice and a reduction in medical errors. However, many medical algorithms are not used because few practitioners have the time to locate useful medical algorithms, let alone convert these from paper to computational formats. For these reasons the authors decided to create a searchable knowledge-base containing computable versions of paper-based medical algorithms. Clearly, there are many benefits to computer-based algorithms over manual execution using a calculator. These include error-free execution every time (provided the algorithm is programmed correctly!) especially for more complex formulae that include mathematical functions like exponentiation.

### *Implementation into Spreadsheets*

This decision led to a search for the most suitable computing technology. After some discussion Excel spreadsheets were selected. There are several advantages to using spreadsheets:

1. Due to the dominance of Microsoft Office, spreadsheets are ubiquitous and available even on many mobile devices
2. The spreadsheet interface is familiar to many clinicians
3. The rich set of formulas built in to Excel spreadsheets means programming is not needed to encode medical algorithms
4. Spreadsheets are relatively simple to create, modify and maintain.
5. Helps and hints for using the algorithm can be easily added into the spreadsheet itself. However, detailed algorithm documentation and references require a separate word document.

The first online version of medal.org was released in 1998 with about 860 spreadsheets. These spreadsheets were divided across 14 chapters corresponding to topics in medicine such as neurology, oncology, gastroenterology and others. From the home page users were presented with a Table of Contents with one hyperlink entry for each chapter. Clicking on such an entry led the user to a page containing a list of algorithms for that chapter. Spreadsheets corresponding to a chapter were collected into a workbook enabling users to download an entire chapter at a time. However, in 2001 workbooks became impractical and unwieldy due to the rapid increase in the number of algorithms in each chapter as well as the increase in the number of chapters. For this reason the workbook format was replaced by one in which clicking on the name of an algorithm in a chapter enabled the user to either download the corresponding spreadsheet or execute it on line. In 2008, a custom program was used to convert the spreadsheets into ASPX pages; this preserved the spreadsheet metaphor and the ability to execute the spreadsheet as an ASPX (Active Server pages) server-side application.

An interesting use of Excel was that from 1998 to 2002 the entire HTML code for the medal.org web site was created by writing an HTML code generator in VBA (Visual Basic for Applications) for Excel. To achieve this a master spreadsheet was created that contained the names of all the algorithms and their chapter names, numbers and the file names of the sheets . This master sheet served as an index into a database of spreadsheets. The code generator was written in VBA and included as a macro in this master spreadsheet. For each version (a new version was created about once every 6 months) after populating the master spreadsheet with the information about new algorithms and spreadsheets the HTML code for the entire web site could be created and written out to disk by running this macro.

## Current Status

The Medical Algorithms Project has been available for 11 years. Access is free, but as of September, 2004, new users must register for access. The material is directly linked (non-login access) from computers in the United States Veterans Administration (VA) healthcare system.

The site has about 1,600 unique visits every day and is the #1 result returned from a web search for the terms 'medical algorithms' or 'algorithmic medicine'.

## Profiles of registered users

As of March 2009, the site has about 106,900 registered users from over 200 countries. The majority of registered users are in the USA and the UK. Apart from papers about this resource at various venues such as AMIA (American Medical Informatics Association), no marketing has ever been done. Word of mouth and web search for specific topics are the typical ways in which the web site is found.

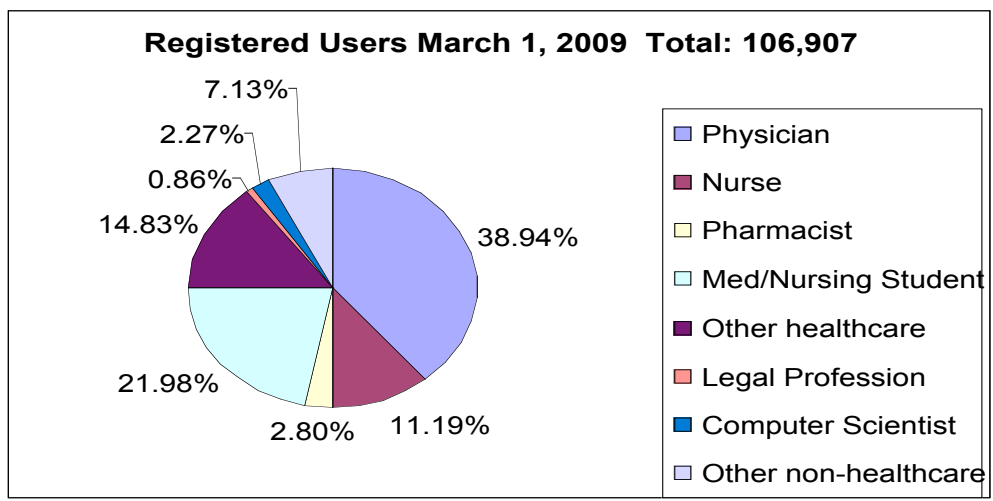

Figure 2: Distribution of registered users by profession

## Challenges

As can be imagined in an ever-growing web site that currently includes 13,500 sheets, a number of challenges have been encountered over the years. Most of these have related to the specific algorithms such as the occasional correction to a formula or the issuance of a new version by the original creators. The spreadsheet format itself has provided to be robust and enduring; however, it has limitations for providing an enhanced user interface. The 400 algorithms written in PHP do provide these features.

A key concern has always been quality assurance with respect to correctness of formulas, use of outdated algorithm versions, keeping up with current literature and so on. Apart from these, so to speak, editorial issues, the resource has given rise to a nascent are of clinical informatics called *medical computation problems* that is concerned with optimal algorithms for finding the best algorithm(s) to apply in a given clinical context. Some issues in this regard are to match up algorithms and patients. For example, an algorithm designed for a pediatric population may give misleading information when used for an adult.

## *Issues for Future Development*

The Medical Algorithm Projects continues to evolve and expand.

Future development will entail:
(1) a bidirectional interface to the medical record. This would overcome a serious limitation in that the current web site requires manual data entry. Manual data entry can be tedious and a source of preventable error.
(2) tools to improve the user experience, including automated selection
(3) integration into medical guidelines
(4) tools for data mining which can be used to improve processes or provide educational opportunities

It should be noted that, due to liability concerns, the spreadsheets are specifically approved only for teaching and research, not for patient care. Approval and validation to meet the good manufacturing requirements for a medical device as specified by the US FDA (Food and Drug Administration) is required to apply this decision support technology into actual patient care.

A recent development has been interest from publishers like Springer in producing specialized collections of medical algorithms in printed form. Springer has started issuing the Medal series on Medical Algorithms. The first one in the series is on Rheumatology and this has been issued in English, German, and Spanish.

## *Conclusions*

Medical algorithms are a valuable but underutilized resource in healthcare. Their use by clinicians and medical researchers, especially at the point of care, can significantly improve the quality and cost-effectiveness of medical care. The Medical Algorithms Project encodes more than 13,500 medical algorithms for 45 areas of medicine into Excel spreadsheets and makes these available freely at the web site www.medal.org.